\RequirePackage{fix-cm}
\documentclass[pdflatex,mathphys]{sty}

\usepackage[draft,nomargin,inline,author=]{fixme}
\usepackage{graphicx}
\usepackage{lmodern}
\usepackage{caption}
\usepackage{enumitem}
\usepackage{textgreek}
\usepackage{amsmath}
\usepackage{amssymb}
\usepackage[capitalize]{cleveref}
\usepackage{mathrsfs}
\DeclareFontShape{U}{rsfs}{m}{n}{ <-7> rsfs5    <7-8> rsfs7    <8-> rsfs10 }{}
\fxsetup{theme=color,mode=multiuser}
\definecolor{fxwarning}{rgb}{1.0,1.0,1.0}
\definecolor{electricviolet}{rgb}{0.56, 0.0, 1.0}
\definecolor{forestgreen}{rgb}{0.13, 0.55, 0.13}

\FXRegisterAuthor{tfl}{atfl}{tfl}

\usepackage{upgreek}
\usepackage[T1]{fontenc}

\jyear{2024}%

\begin{document}

\title[Online Training and Pruning]{Online Training and Pruning of Multi-Wavelength Photonic Neural Networks} 

\author*[1,4]{\fnm{Jiawei} \sur{Zhang}}\email{jiawei.zhang@princeton.edu}
\author[1,4]{\fnm{Weipeng} \sur{Zhang}}\email{weipengz@princeton.edu}
\author[2,4]{\fnm{Tengji} \sur{Xu}}\email{tengjixu@link.cuhk.edu.hk}
\author[1]{\fnm{Lei} \sur{Xu}}\email{leixu@princeton.edu}
\author[1]{\fnm{Eli A.} \sur{Doris}}\email{edoris@princeton.edu}
\author[3]{\fnm{Bhavin J.} \sur{Shastri}}\email{shastri@ieee.org}
\author[2]{\fnm{Chaoran} \sur{Huang}}\email{crhuang@ee.cuhk.edu.hk}
\author*[1]{\fnm{Paul R.} \sur{Prucnal}}\email{prucnal@princeton.edu}

\affil*[1]{\orgdiv{Department of Electrical and Computer Engineering}, \orgname{Princeton University}, \orgaddress{\city{Princeton}, \postcode{08544}, \state{New Jersey}, \country{USA}}}

\affil[2]{\orgdiv{Department of Electronic Engineering}, \orgname{The Chinese University of Hong Kong}, \orgaddress{\city{Shatin}, \state{Hong Kong SAR}, \country{China}}}

\affil[3]{\orgdiv{Department of Physics, Engineering Physics and Astronomy}, \orgname{Queen’s University}, \orgaddress{\city{Kingston}, \postcode{K7L 3N6}, \state{Ontario}, \country{Canada}}}

\affil[4]{The authors contribute equally to this paper}

\abstract{CMOS-compatible photonic integrated circuits (PICs) are emerging as a promising platform in artificial intelligence (AI) computing. Owing to the compact footprint of microring resonators (MRRs) and the enhanced interconnect efficiency enabled by wavelength division multiplexing (WDM), MRR-based photonic neural networks (PNNs) are particularly promising for large-scale integration. However, the scalability and energy efficiency of such systems are fundamentally limited by the MRR resonance wavelength variations induced by fabrication process variations (FPVs) and environmental fluctuations. Existing solutions use post-fabrication approaches or thermo-optic tuning, incurring high control power and additional process complexity. In this work, we introduce an online training and pruning method that addresses this challenge, adapting to FPV-induced and thermally induced shifts in MRR resonance wavelength. By incorporating a power-aware pruning term into the conventional loss function, our approach simultaneously optimizes the PNN accuracy and the total power consumption for MRR tuning. In proof-of-concept on-chip experiments on the Iris dataset, our system PNNs can adaptively train to maintain a 96\% classification accuracy, while achieving a 44.7\% reduction in tuning power via pruning. Additionally, our approach reduces the power consumption by orders-of-magnitude on larger datasets. By addressing chip-to-chip variation and minimizing power requirements, our approach significantly improves the scalability and energy efficiency of MRR-based integrated analog photonic processors, paving the way for large-scale PICs to enable versatile applications including neural networks, photonic switching, LiDAR, and radio-frequency beamforming.
}


\maketitle
\section{Introduction}
\label{sec:Intro}

Large neural networks (NNs) have demonstrated exceptional performance in edge computing \cite{chen2019deep}, natural language processing \cite{ouyang2022training}, and autonomous systems \cite{lecun2015deep}. CMOS-compatible silicon photonic integrated circuits (PICs) are emerging as a promising platform in artificial intelligence (AI) computing \cite{shen2017deep,shastri2021photonics}, offering significant advantages including low latency, high bandwidth, and fully programmability \cite{shastri2021photonics,bandyopadhyay2024single,chen2023all}. Integrated photonic neural networks (PNNs) generally fall into two major categories: coherent PNNs based on interferometric meshes (e.g., Mach-Zehnder interferometers (MZIs)) \cite{shen2017deep,bandyopadhyay2024single} and wavelength-division multiplexed (WDM) PNNs based on wavelength-selective filters (e.g., microring resonators (MRRs)) \cite{tait2017neuromorphic,feldmann2021parallel}. Owing to the compact footprint of MRRs and the enhanced interconnect efficiency enabled by WDM, MRR-based PNNs can be implemented with significantly less chip area than their coherent equivalents \cite{lederman2024low}, and are promising for large-scale integration using CMOS-compatible silicon photonic foundry processes \cite{siew2021review,giewont2019300,chrostowski2019silicon}. In addition to NN inference \cite{shastri2021photonics,huang2022prospects}, such MRR-based integrated analog photonic processors have also found important applications in photonic switching \cite{jayatilleka2019photoconductive}, LiDAR \cite{larocque2019beam,huang2025integrated}, RF beamforming \cite{nichols2024photonics,liu2018ultra}, and data interconnects \cite{rizzo2023massively,daudlin2025three}.

However, a key challenge in realizing large-scale MRR-based analog photonic processors is the functional variation of MRRs caused by unavoidable fabrication process variations (FPVs) and dynamic environmental fluctuations (e.g., thermal crosstalk and polarization drifts), which can induce significant random shifts in the MRR resonance wavelengths. This shift can be expressed as \cite{bogaerts2012silicon}:
\begin{equation}
\label{Equation: FPV_resonance shift}
\delta \lambda_{0}=\frac{\delta^{(\mathrm{env})} n_{\mathrm{eff}} \cdot \lambda_{0}}{n_g},
\end{equation}
where $\lambda_0$ is the MRR resonance wavelength, $\delta^{(\mathrm{env})} n_{\mathrm{eff}}$ is the effective index shift due to environmental changes, and $n_g$ is the group index accounting for waveguide dispersion. The effective index shift can be further expressed as \cite{densmore2006silicon}
\begin{equation}
\delta^{(\mathrm{env})} n_{\mathrm{eff}}=c \int \Delta \varepsilon \mathbf{E}_v \cdot \mathbf{E}_v^* d x d y
\end{equation}
where $c$ is the speed of light, $\Delta \varepsilon (x,y)$ denotes a local change in the  dielectric constant, and $\mathbf{E}_v (x,y)$ is the normalized modal electric field vector of the waveguide mode. All the static and dynamic variations---including sub-wavelength FPVs in geometric parameters, ambient temperature disturbances, and thermal crosstalk---can alter $\Delta \varepsilon (x,y)$, resulting in a resonance shift $\delta \lambda_{0}$ comparable to the free spectral range (FSR) of MRRs \cite{tait2022quantifying}. For example, in a recent study \cite{chrostowski2014impact}, measurements of 371 identically designed racetrack-shaped resonators, revealed resonance shifts ranging from 1.76 nm (median) to 6 nm (maximum), as a result of inherent silicon thickness variations ($\pm$5 nm fluctuations in 220 nm layers) across wafers and fabrication batches. It is also shown that fluctuations in ambient temperature can lead to a drift in the MRR resonance wavelength of tens of pm, resulting in a degradation of the accuracy of an MRR-based photonic neural network (PNN) to 67\% from 99\% for a two-layer MNIST classification \cite{xu2024control}. While previous work has explored various post-fabrication strategies to counteract these variations---including Germanium (Ge) ion implementation \cite{milosevic2018ion,chen2018real}, integration of phase change materials (PCMs) \cite{rios2015integrated,fang2022ultra,zheng2018gst}, and deposition of photochromic materials \cite{bilodeau2024all,xu2025building}---these techniques are only effective at correcting static FPVs, and they require additional post-fabrication processing complexity and precise control over the materials involved. Alternatively, thermo-optic tuning remains widely used method due to its broad tuning range, but it is power-intensive; consuming 28 mW/FSR using an embedded N-doped heater \cite{jayatilleka2015wavelength, huang2020demonstration}. Consequently, the scalability and energy efficiency of MRR-based analog photonic processors are fundamentally constrained, hindering their applications to large-scale networks such as large AI models and WDM transceivers. 

To address fundamental limitation, we propose online training and pruning in MRR-based PNNs that adapt to FPV-induced and thermally induced shifts in MRR resonance wavelength. Based on perturbation-based gradient descent algorithm, we develop an online training framework that maps the trainable NN parameters to MRR-based PNN chips without the need of LUTs. We further incorporate a power-aware pruning term into the conventional loss function, which simultaneously optimizes the PNN accuracy and the total power consumption for MRR tuning. In proof-of-concept on-chip experiments, we demonstrate online training with an iterative feedback system with a PIC performing fast analog matrix-vector multiplications (MVMs), combined with a central processing unit (CPU) digitally computing gradients at a slower timescale. Using a 3×2 PNN on the Iris dataset, our system PNNs can adaptively train to maintain a 96\% classification accuracy across static (FPV) or dynamic (thermal drifts) variations, while achieving a 44.7\% reduction in tuning power via pruning. Additionally, simulations with larger and deeper convolutional neutral networks (CNNs) on standard datasets---including MNIST \cite{lecun1998gradient}, CIFAR-10 \cite{krizhevsky2009learning}, and Fashion-MNIST \cite{xiao2017fashion}--- validate the scalability of our method, showing orders-of-magnitude reductions in power consumption. By addressing chip-to-chip variation and minimizing power requirements, our approach significantly improves the scalability and energy efficiency of MRR-based integrated analog photonic processors, paving the way for large-scale PICs to enable versatile applications including NNs, photonic switching, LiDAR, and RF beamforming.

\section{Results}
\label{sec:Results}
\subsection{Concept and Principle}
\label{sec:Concept}

\begin{figure}[ht!]
\centering
\includegraphics[width=.99\linewidth]{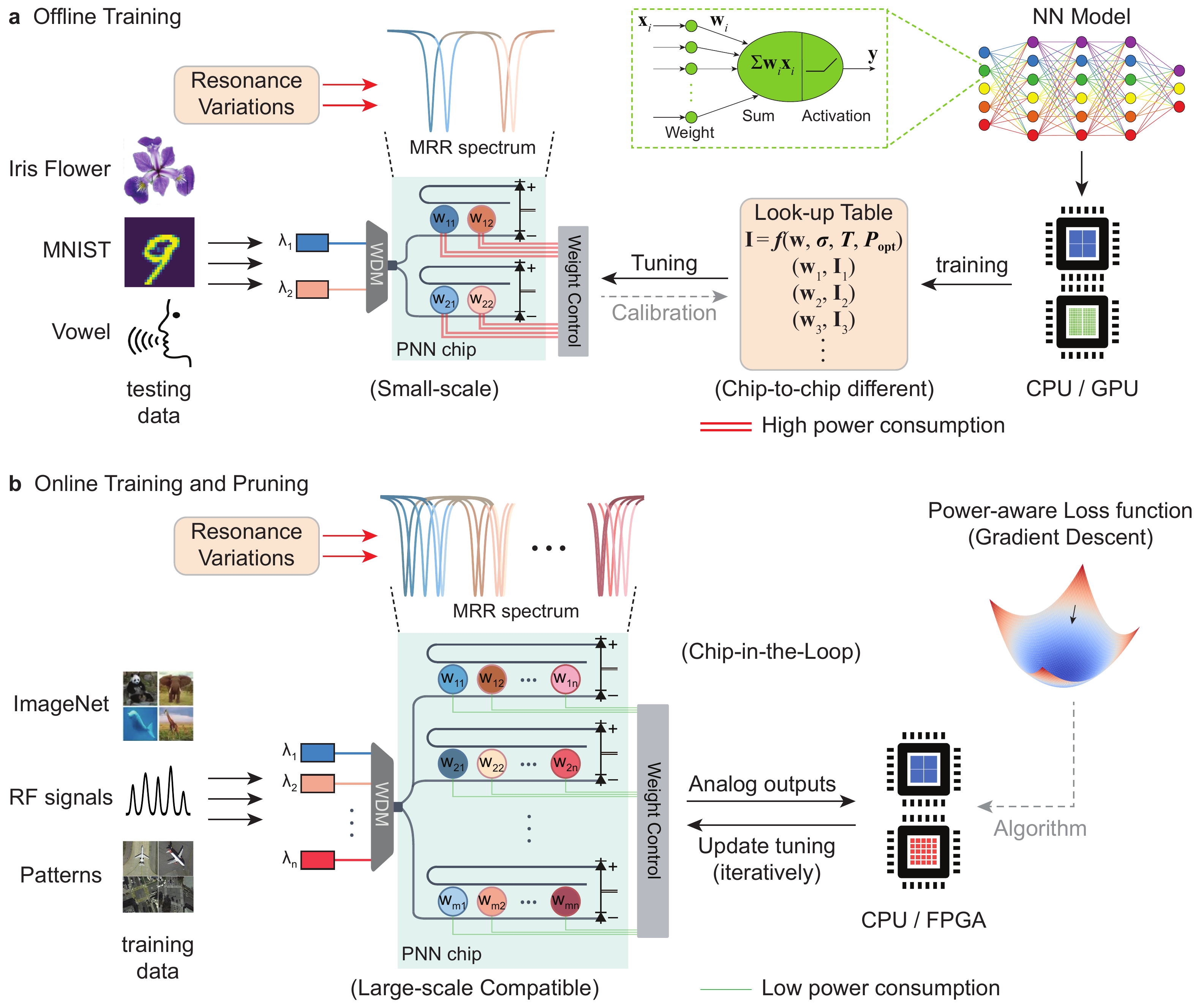}
\caption{
\textbf{a} Conventional offline training method supporting only small-scale PNN chips. In this approach, NN parameters are calculated in software and mapped onto the PNN chips. The resonance variations of MRRs necessitate complicated look-up tables, and lead to higher power consumption for MRR control. \textbf{b} Online training and pruning method compatible for large-scale PNN chips. The training of PNNs occurs on the same chip used for inference, accounting for any chip-to-chip variations and environmental fluctuations. Our approach simultaneously optimizes the PNN accuracy and the total power consumption for tuning all the MRRs. 
}
\label{fig:vision_figure}
\end{figure}

\subsubsection{Offline Training}
\label{sec:Offline Training}

While PNNs can operate with picosecond latency for real-time applications \cite{huang2021silicon,zhang2024system,zhou2024silicon}, they rely on slower digital computers (i.e. CPU or GPU) for training, a process called offline training \cite{huang2021silicon,lin2018all}. As shown in Fig. \ref{fig:vision_figure}a, in this approach, NN parameters (e.g., neuron weights and biases) are optimized in software using backpropagation (BP) based gradient descent algorithm:
\begin{equation}
\label{BP offline}
\textbf{w}^{k+1} \leftarrow \textbf{w}^k-\alpha \nabla_{\textbf{w}^k} \mathcal{L},
\end{equation}
where $\textbf{w}^k$ denotes the weight matrix in the $k$th training epoch, $\alpha$ is the learning rate, $\mathcal{L}$ is the loss function. The parameters are then mapped onto the physical hardware (PNN chips) using highly accurate calibration models (also called look-up table (LUT)). For the MRR in the $i$th row, the $j$th column of an $M\times N$ MRR weight bank, it writes as 
\begin{equation}
\label{look-up table}
    \textbf{I}_{ij} = f_{ij}(\textbf{w}_{ij}) = f_j(\textbf{w}_{ij}).
\end{equation}
Here $\textbf{w}_{ij}$ denotes the weight to be executed on the $i,j$th MRR, and $\textbf{I}_{ij}$ is the required tuning current. The LUT, denoted as a mapping function $f_{ij}$, is determined by the relative position of the $j$th laser wavelength and the MRR resonant state. Ideally, the MRRs operating at the same laser wavelength are expected to exhibit identical resonant wavelengths, simplifing the LUT to $f_j$.

However, in practice, the MRRs in the same column may exhibit significantly different resonant wavelengths due to static and dynamic variations. As derived in Supplementary Note 1, an LUT for the $i,j$th MRR---accounting for FPVs, ambient temperature variations, input optical power, and the self heating effect---is approximately given by
\begin{equation}
\label{Equation: look-up table real}
    \textbf{I}_{ij} = \tilde{f}_{ij}(\textbf{w}_{ij}, \sigma_{ij}, T, P_{\text{opt}}),
\end{equation}
where $\sigma_{ij}$ denotes the deviation of resonant wavelength due to FPVs, $T$ is the time-varying ambient temperature, $P_{\text{opt}}$ is the input optical power. Based on Eq. (\ref{Equation: look-up table real}), it remains complicated to generate an LUT that accounts for all the non-idealities across large-scale MRR weight banks. Any inaccuracies in LUTs will directly introduce arbitrary deviations in the mapped weight parameters, leading to significant performance degradation across NN system benchmarks such as classification tasks.

In an MRR-based weight bank using thermo-optic tuning to counteract these non-idealities, the required electrical power for actively programming weights of the $i,j$th MRR can be written as \cite{tait2022quantifying}: 
\begin{equation}
\label{Equation: Power of single MRR}
P^{\text{weight}}_{i j}=\textbf{I}_{i j}^2 R=P_{i j}^{\text{lock}}+P_{i j}^{\text {conf}},
\end{equation}
where $R$ is the resistance of the metal heater. The power breaks down into a static weight locking power that locks the MRR resonance to the desired wavelength, and a configuration power to program the weight value \cite{tait2022quantifying}. Therefore, the total power for the weight configuration consumed by an $M\times N$ MRR weight bank is
\begin{equation}
\label{Equation: total power offline}
    \textbf{P}^{\text{weight}} = \Sigma_{i,j} P^{\text{weight}}_{ij}.
\end{equation}
For simplicity, we denote $\textbf{P}^{\text{weight}}$ as $\textbf{P}$ in the following sections except Section~\ref{sec:Large Scale}.

\subsubsection{Online Training}
\label{sec:OnlineTraining}

Online training, also known as "in situ" or "chip-in-the-loop" training, was originally proposed to alleviate the intensive use required of CPU/GPUs for training non Von-Neumann architecture based computing hardware
\cite{buckley2023photonic}, where the training occurs on the same hardware used for inference. It was also recognized that online training can mitigate hardware nonidealities, and promise for iterative weight updates in real-time \cite{filipovich2022silicon}. The state-of-the-art online training algorithms for PNNs \cite{hughes2018training,filipovich2022silicon,bandyopadhyay2022single,zhang2021efficient,pai2023experimentally} can be classified into two major categories: gradient-based and gradient-free algorithms. Gradient-free algorithms, such as genetic algorithms \cite{zhang2021efficient}, are straightforward to implement in practice but often face inherent challenges with convergence and scalability. In contrast, gradient-based algorithms are typically more efficient and, therefore, more widely adopted. The most commonly used gradient-based training algorithm for software-based NNs, backpropagation, analytically computes gradients by back-propagating errors using the chain rule \cite{lecun1998gradient}. While significant progress has been made in experimentally realizing BP on photonic hardware \cite{pai2023experimentally, hermans2015trainable, gu2022training}, the process typically requires global optical power monitoring and evaluation of nonlinear activation function gradients in software. This approach introduces additional system complexity and latency overhead. Other gradient-based algorithms, such as direct feedback alignment \cite{filipovich2022silicon}, replace the chain rule in back-propagation with a random weight matrix, but their validation has been limited to small, shallow NNs. 

In our approach, the online training of our MRR-based PNNs is implemented by a perturbation-based gradient descent algorithm, which estimates gradients only based on forward inference running on PNN chips. Here, the NN parameters to be optimized are set to be MRR tuning currents instead of neuron weights:
\begin{equation}
\label{online training}
\textbf{I}^{k+1} \leftarrow \textbf{I}^k-\alpha \nabla_{\textbf{I}^k} \mathcal{L}.
\end{equation}
The gradient of loss function with respect to MRR tuning currents ($\textbf{I}^k$) is approximately given by perturbation measurement:
\begin{equation}
\label{Equation: Perturbation}
\nabla_{\textbf{I}^k} \mathcal{L} \approx \frac{\Delta \mathcal{L}}{\Delta \textbf{I}^k},
\end{equation}
where $\Delta \textbf{I}^k$ is the perturbation rate of the MRR tuning currents in the $k$th training epoch, and $\Delta \mathcal{L}$ is the measured change of loss function induced by the perturbation. In practice, $\Delta \textbf{I}^k$ should be sufficiently small to perserve the validity of approximation in Eq. (\ref{Equation: Perturbation}), while remaining large enough to produce measurable $\Delta \mathcal{L}$ above experimental noise. 

\subsubsection{Pruning}
\label{sec:Pruning}

In software-based NNs, pruning is a model compression technique that removes redundant weight parameters (by setting their values to zero) whilst maintaining accuracy. While the implementation of pruning has also been investigated in PNNs, prior approaches either lack robustness against nonidealities \cite{banerjee2023pruning,gu2021efficient} or require extensive offline training \cite{xu2024control}. As shown in Fig. \ref{fig:vision_figure}b, our proposed approach accounts for the total MRR tuning power in the “chip-in-the-loop” process, by incorporating a power-aware “pruning” term into the conventional loss function:
\begin{equation}
    \label{Equation: pruned loss}
    \tilde{\mathcal{L}} = \mathcal{L} + \gamma \textbf{P}
\end{equation}
where $\tilde{\mathcal{L}}$ is the modified loss function given by the sum of the conventional loss function $\mathcal{L}$ and the power-aware pruning term $\gamma \textbf{P}$, $\gamma$ is an empirically determined hyperparameter that defines pruning strength. According to Eqs. (\ref{Equation: Power of single MRR}) and 
(\ref{Equation: total power offline}), $\tilde{\mathcal{L}}$ can also be expressed as 
\begin{equation}
    \tilde{\mathcal{L}} = \mathcal{L} + \gamma \Sigma_{i,j} P_{ij} = \mathcal{L}(\textbf{I}) + \gamma \Sigma_{i,j} \textbf{I}_{i j}^2 R.
\label{Equation: Online Training with pruning}
\end{equation}
The selection of the hyperparameter $\gamma$ is critical in this framework, as it explicitly parameterizes the trade-off between the NN accuracy and energy efficiency. Depending on the demand of the end users, minimizing this modified loss function can enable concurrent optimization of both the NN accuracy and the total power consumption. For example, in a power-constrained environment, the training problem becomes 
\begin{equation}
\min \mathcal{L}+\gamma \mathbf{P} \quad \text {s.t. } \mathbf{P} \leq \mathbf{P}_{\max }
\end{equation}
with a relatively larger value of $\gamma$, where $\mathbf{P}_{\max }$ denotes the maximum power available for weight configurations. Oppositely, in scenarios prioritizing high accuracies, it becomes 
\begin{equation}
\min \mathcal{L}+\gamma \mathbf{P} \quad \text {s.t. } \mathcal{L} \leq \mathcal{L}_{\max }
\end{equation}
with a relatively smaller value of $\gamma$, where $\mathcal{L}_{\max }$ denotes the upper bound of the conventional loss function permitted under a given accuracy constraint.

\subsection{Experiment}
\label{sec:Experiment}

\subsubsection{Setup}
\label{sec:Setup}
\begin{figure}[ht!]
\centering
\includegraphics[width=.99\linewidth]{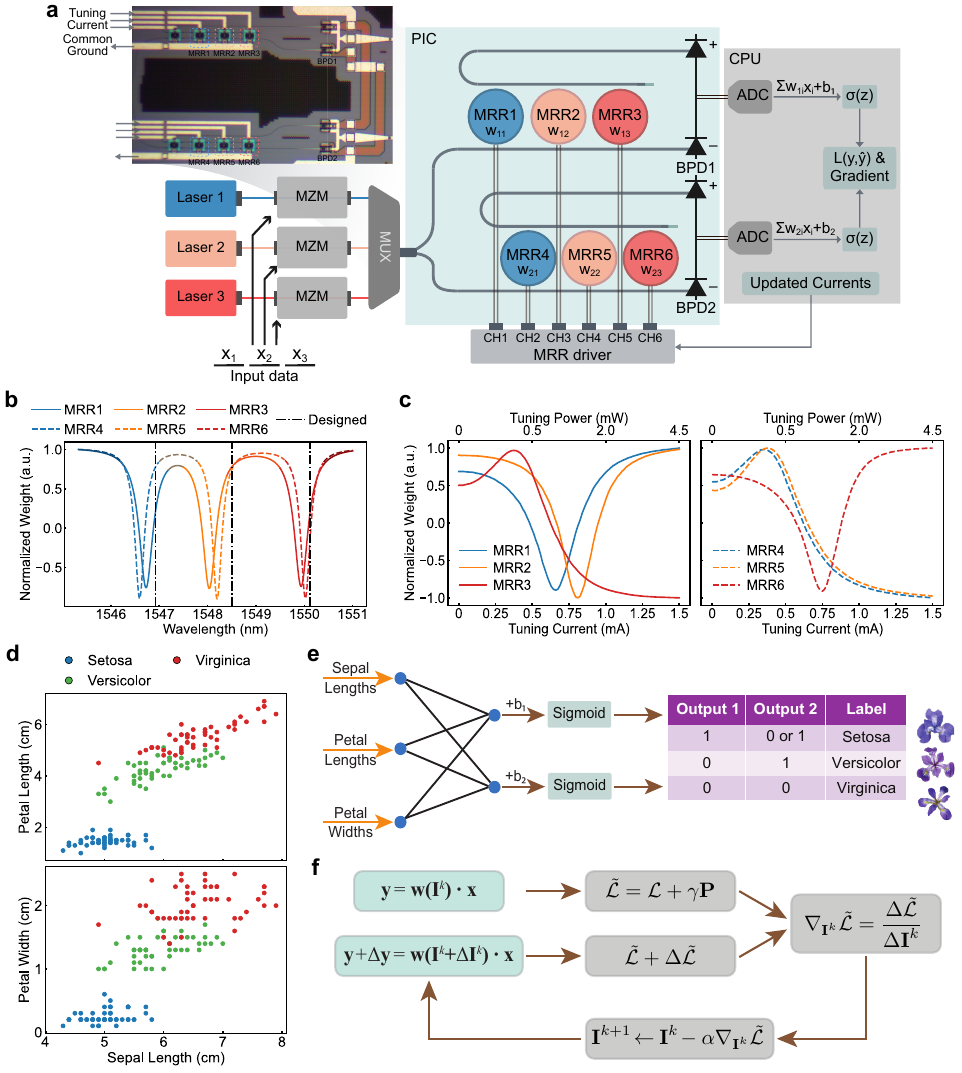}
\caption{\textbf{a} Schematic of our experimental setup. MZM, Mach-Zehnder Modulator. MUX, wavelength multiplexer. PIC, photonic integrated circuit. MRR, microring resonator. BPD, balanced photodetector. ADC, analog-to-digital converter. CPU, central processing unit. $\sigma(z)$ denotes the nonlinear activation function performed on software. The inset shows the micrograph of the two MRR weight banks and BPDs. \textbf{b} Normalized weight spectrum of two MRR weight banks at 20$^\circ$C. The MRRs at the same colors (i.e., MRR1 and MRR4, MRR2 and MRR5, MRR3 and MRR6) are designed to share the same resonance wavelengths, aligned with 200 GHz spaced ITU grids (1546.92 nm, 1548.51 nm, 1550.12 nm). \textbf{c} Tuning characteristics of six MRRs at 20$^\circ$C. \textbf{d} Scatterplot of Iris flower dataset. \textbf{e} Schematic of a simple 3$\times$2 neural network for Iris classification. \textbf{f} Online training and pruning procedure. At each iteration, the PIC performs matrix-vector multiplication without and with perturbations, while the CPU evaluates the gradients of power-aware loss function, and updates the MRR tuning currents for the next iteration. }
\label{fig:schematic}
\end{figure}

Recently, integrated PNNs have been extensively investigated and experimentally validated across various on-chip scales, achieving low latencies of hundreds of picoseconds at diverse applications \cite{ashtiani2022chip,bandyopadhyay2024single,zhang2024system}. Notable demonstrations include a 3-layer PNN with 9 neurons for image classification \cite{ashtiani2022chip}, a 3-layer 6$\times$6 PNN for vowel classification task \cite{bandyopadhyay2024single}, and a single-layer 4$\times$2 PNN for fiber nonlinearity compensation \cite{huang2021silicon}. In our proof-of-concept experiment, a 3$\times$2 MRR weight bank is used to demonstrate our proposed online training and pruning approach and the associated energy savings.

Our experimental setup is illustrated in Fig. \ref{fig:schematic}a. First, three channels of input data (denoted as $\textbf{x}_1$, $\textbf{x}_2$, $\textbf{x}_3$) are generated from a high sampling rate signal generator (Keysight M8196A), and modulated onto laser 1, 2, and 3 (Pure-Photonics, PPCL500) respectively via Mach-Zehnder modulators (MZMs). The lights from these three lasers, each at a distinct wavelength, are combined using a wavelength-division multiplexer (MUX) and then split equally between two MRR weight banks. Each MRR weight bank consists of three MRRs, which are designed with slightly different radii (8 \textmu m, 8.012 \textmu m, 8.024 \textmu m). The MRRs at the same column (i.e., MRR1 and MRR4, MRR2 and MRR5, MRR3 and MRR6) are expected to share the same corresponding resonance wavelengths, aligned with 200 GHz spaced ITU grids (1546.92 nm, 1548.51 nm, 1550.12 nm, respectively). However, as shown in Fig. \ref{fig:schematic}b, the measured weight bank spectrum indicates that the resonance wavelengths of all six MRRs deviate from the designed values due to FPVs and temperature changes. To counteract FPVs, the wavelengths of three lasers are further tuned (1546.88 nm, 1548.24 nm, and 1550.05 nm) to align with the deviated resonance wavelengths. Moreover, all MRRs are thermally tuned via embedded N-doped heaters to actively program and configure weights (denoted as $\textbf{w}_{11}$, $\textbf{w}_{12}$, $\textbf{w}_{13}$, $\textbf{w}_{21}$, $\textbf{w}_{22}$, $\textbf{w}_{23}$), allowing for individual weighting of input analog data in the three wavelength channels. The thermal tuning characteristics of six MRRs at 20$^{\circ}$C, are illustrated in Fig. \ref{fig:schematic}c.  

Our proof-of-concept experiment is validated on a standard Iris dataset, where three species of Iris flowers (Setosa, Versicolour, Virginica) are classified using only three out of four input features (sepal length, petal length, and petal width), as shown in Fig. \ref{fig:schematic}d. This three-class classification problem is converted into a two-step binary classification (Fig. \ref{fig:schematic}e) mapped to the on-chip PNN, which utilizes eight trainable parameters. The parameters include the tuning currents of six MRRs --- $\textbf{I}_{11}$, $\textbf{I}_{12}$, $\textbf{I}_{13}$, $\textbf{I}_{21}$, $\textbf{I}_{22}$, $\textbf{I}_{23}$ (corresponding to six weights) --- and two biases, $b_1$ and $b_2$. The 150 samples are split into 120 for training and 30 for testing. 

The analog optical signals from the drop and through port are then captured and differentiated by two balanced photodetectors (BPDs), which gives the electrical output of the weighted summations, $\Sigma_i \textbf{w}_{1i}\textbf{x}_i$ and $\Sigma_i \textbf{w}_{2i}\textbf{x}_i$. The output is further read by an oscilloscope and demodulated in a CPU, which evaluates the power-aware loss function (Eq. (\ref{Equation: pruned loss})) and calculates gradients based on perturbation (Eq. (\ref{Equation: Perturbation})). Finally, the CPU updates the MRR tuning currents for the next training epoch (Eq. (\ref{online training})) and commands the MRR driver, which is equipped on a customized printed circuit board (PCB) (Supplementary Note 2). As shown in Fig. \ref{fig:schematic}f, the training of our PNN occurs with the photonic chip in the loop, iteratively optimizing the loss function and reducing the total power consumption. 

\subsubsection{Demonstration on 3$\times$2 PNN}
\label{sec:Demonstration}

\begin{figure}[ht!]
\centering
\includegraphics[width=.86\linewidth]{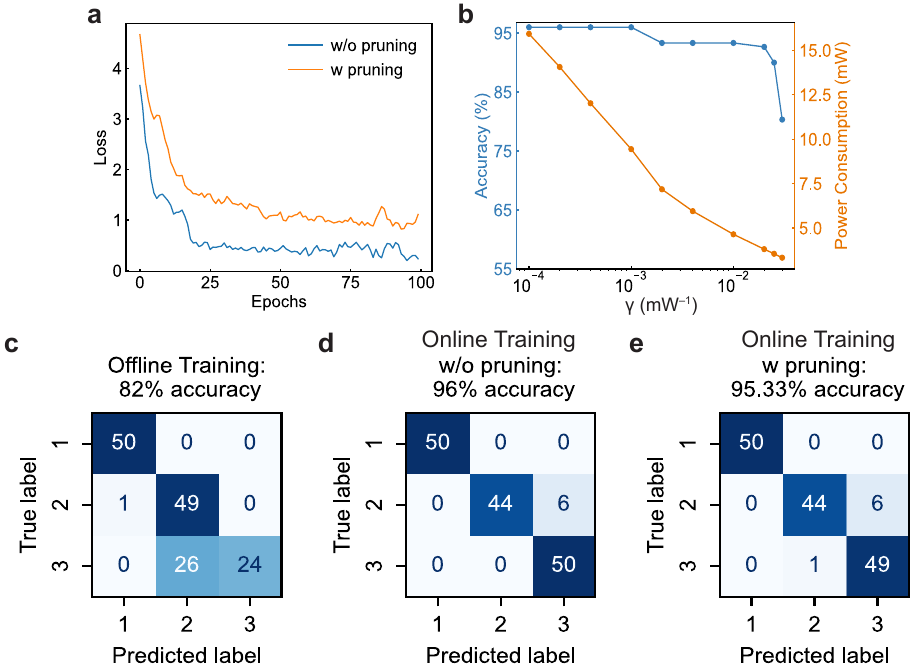}
\caption{\textbf{a} Experimental results of online training losses without, or with the pruning method. \textbf{b} Simulated result indicating the tradeoff between the prediction accuracy and power efficiency. \textbf{c–e} Confusion matrices for the 150 samples, obtained by the conventional offline training, online training without, or with the pruning method, respectively.}
\label{fig:Iris results with pruning}
\end{figure}

PNN training is conducted at 20$^\circ$C under three conditions: conventional offline training, online training without pruning, and online training with pruning. Each training configuration is repeated 10 times to ensure consistency. The perturbation rate of MRR tuning currents ($\Delta\textbf{I}^k$) is set to 0.05 mA. To evaluate the gradients with respect to the MRR tuning currents ($\nabla_{\textbf{I}^k} \mathcal{L}$), each training epoch batches the training data seven times through the PNN: six times perturbing each tuning current ($\textbf{I}^k_{ij} + \Delta \textbf{I}^k_{ij}$) and one time to measure $\tilde{\mathcal{L}}$ with non-perturbed currents $\textbf{I}^k_{ij}$. 
The plots in Fig. \ref{fig:Iris results with pruning}a show the average of both categorical cross-entropy loss ($\mathcal{L}$) and power-aware loss ($\tilde{\mathcal{L}}$) versus training epoch. It is observed that the PNN can quickly converge to the optimal weights within 25 epochs using perturbation-based online training.

We also simulate the trade-off between the NN accuracy and the associated energy savings (Fig. \ref{fig:Iris results with pruning}b). The simulation indicates that the trade-off is optimized at $\gamma_{\mathrm{opt}}=0.01$ mW$^{-1}$, where significant reduction of overall tuning power (exceeding 70\%) can be achieved before the accuracy drops off. As predicted in Section \ref{sec:Pruning}, in scenarios prioritizing high accuracies where $\gamma<\gamma_{\mathrm{opt}}$, the loss function $\tilde{\mathcal{L}}$ is dominated by the conventional loss function $\mathcal{L}$, such that the PNN maintains a high accuracy of >93\%. Contrarily, in scenarios constraining overall power consumption where $\gamma>\gamma_{\mathrm{opt}}$, the power-aware loss function $\tilde{\mathcal{L}}$ is dominated by the pruning term $\gamma \textbf{P}$, leading to a larger amount of energy savings but may degrade the PNN accuracy. As shown by the confusion matrix in Fig. \ref{fig:Iris results with pruning}c, the offline training experiment only produces 82\% accuracy on the 150 samples. In the accuracy-prioritized online training experiment (without pruning, $\gamma$ set to 0) (Fig. \ref{fig:Iris results with pruning}d), the overall classification of the Iris flower task is improved to 96\%, with a total MRR tuning power of 9.54 mW. In the online training experiment constraining power consumption (with pruning, $\gamma$ set to 0.0075 mW$^{-1}$) (Fig. \ref{fig:Iris results with pruning}e), we observe a 44.7\% reduction (from 9.54 mW to 5.28 mW) of total tuning power while the PNN maintains a classification accuracy of 95.33\%. 

\begin{figure}[ht!]
\centering
\includegraphics[width=.86\linewidth]{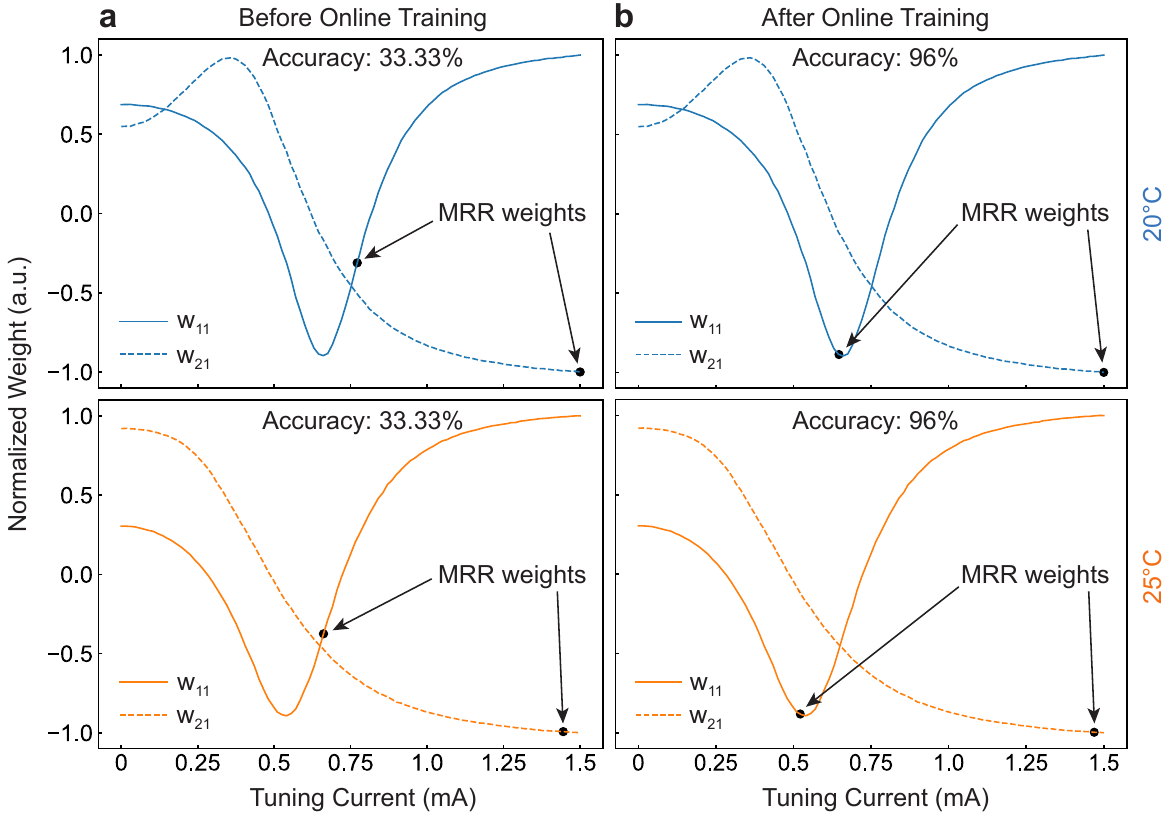}
\caption{Adaptive training at 20$^{\circ}$C (the upper plots) and 25$^{\circ}$C (the lower plots). \textbf{a, b} illustrate two of the six MRR tuning currents (weights $\textbf{I}_{11}$ and $\textbf{I}_{21}$) before and after online training, respectively.}
\label{fig:Iris results with temperature}
\end{figure}

Furthermore, to demonstrate the adaptability of our online training method to temperature drifts, additional experiments are validated at two different temperatures (20$^{\circ}$C and 25$^{\circ}$C). Before online training, the MRR tuning currents are randomly initialized within the tuning range of 0 – 1.5 mA (only weights $\textbf{w}_{11}$ and $\textbf{w}_{21}$ are plotted, Fig. \ref{fig:Iris results with temperature}a), producing untrained classification outputs with only 33.33\% accuracy. After online training for 100 epochs (Fig. \ref{fig:Iris results with temperature}b), the PNN can be trained to the optimal weights, regardless of the changes of MRR tuning characteristics due to temperature drifts. As a result, high classification accuracy (96\%) is consistently achieved at both temperatures. The proof-of-concept experimental results further validate that our online training and pruning method can handle any PNN hardware non-idealities, including chip-to-chip FPVs, temperature fluctuations, and even nonlinearities induced by input optical power.

\subsection{Energy Savings in large-scale NNs via Pruning}
\label{sec:Large Scale}

\begin{figure}[ht!]
\centering
\includegraphics[width=.7\linewidth]{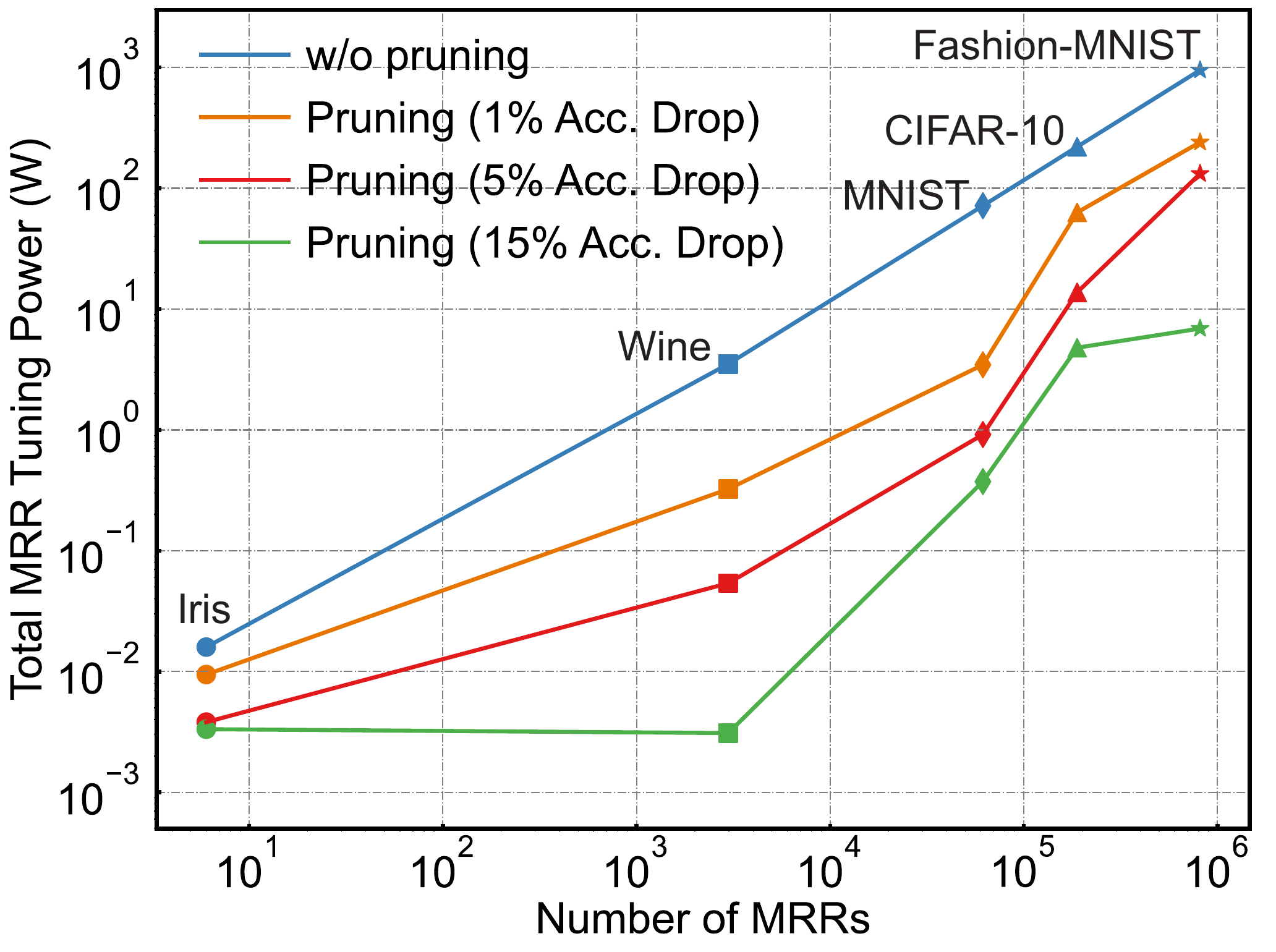}
\caption{Simulated results on reducing overall power consumption enabled by online pruning across different size MRR-based PNNs. }
\label{fig:Scalability with pruning}
\end{figure}

In addition to the MRR tuning power for configuring weights, the overall power consumed by a multiwavelength PNN also includes the power needed for laser pumping and O-E/E-O data conversions (e.g., modulators, photodetectors, and analog-to-digital converters (ADCs)). For a multiwavelength PNN with $N$ neurons and $N^2$ MRRs, the overall power can be expressed by \cite{tait2022quantifying}:
\begin{equation}
\label{Equation: Total PNN power}
\textbf{P}^{\text{total}}=N^2 \times \textbf{P}^{\text {weight}}+N \times \textbf{P}^{\text{laser}}+N \times BW \times \textbf{E}^{\text {OEO}},
\end{equation}
where $\textbf{P}^{\text{laser}}$ is the power needed for laser pumps, $BW$ represents the bandwidth of the signal modulated on the optical carriers, and $\textbf{E}^{\text {OEO}} = \textbf{E}^{\text{mod}} + \textbf{E}^{\text{det}} + \textbf{E}^{\text{ADC}}$ is the energy of O/E/O data conversion, associated with modulation, detection, and ADCs. The MRR tuning power $\textbf{P}^{\text {weight}}$ is shown to scale quadratically with the number of neurons, whereas the power use of laser pumps and data conversions scales linearly. While various technologies have been demonstrated to address different power contributors \cite{timurdogan2014ultralow,bilodeau2024all,xiang2022high}, we specifically look into MRR tuning power dominated scenarios and the energy savings in larger-scale multiwavelength PNNs via online pruning.  

We extend the simulations to larger, deeper feedforward and CNN architectures for various classification tasks with standard datasets, including scikit-learn Wine dataset, MNIST, CIFAR-10, and Fashion-MNIST. The FPVs in the MRRs are simulated by inducing Gaussian-distributed variations (with a standard deviation of $\sigma$) to their resonance wavelengths (Supplementary Note 4) \cite{tait2022quantifying}. We quantify the impact of online pruning on inference power consumption, defined as the control power for all the weighting MRRs. Consistent with the theoretical framework in Section \ref{sec:Pruning}, the accuracy-power trade-off for large-scale NNs is governed by the pruning strength $\gamma$, as exemplified by the Iris dataset results in Fig. \ref{fig:Iris results with pruning}b. As shown in Fig. \ref{fig:Scalability with pruning}, the baseline unpruned PNNs (blue line) exhibit a linear increase in total MRR tuning power with the PNN scale (number of MRRs), prioritizing accuracy under unlimited power budgets ($\gamma = 0$). In contrast, for power-constrained systems tolerating specific classification errors, our method reduces power consumption by orders of magnitude. For instance, in the simulation using the scikit-learn Wine dataset (featuring a three-layer feedforward NN with 2976 weights), our method achieves total MRR tuning power reductions of 10.8$\times$, 65.2$\times$, and 1133.4$\times$ under validation accuracy drops of 1\%, 5\%, and 15\%, respectively. These findings significantly improve the scalability and energy efficiency of MRR-based PNNs, particularly paving the way for their operations in power-constrained scenarios such as edge computing, wearable devices, and autonomous driving systems.

\section{Discussion}
\label{sec:Discussion}

With the emergence of new AI computing paradigms, such as brain-inspired neuromorphic computing and quantum computing, online training has been proposed that revolutionizes the definitions of unconventional hardware computing architecture and AI training algorithms. In this work, we focus on online training and pruning for integrated silicon photonic neural networks, which among the new computing paradigms appear attractive for their full-programmability and CMOS-compatibility. Our proposed approach provides a methodology that breaks one fundamental limiting factor on scalability due to chip-to-chip variations, and simultaneously optimizes overall power consumption in thermally controlled MRR-based PICs. 

Recent advances in photonic AI hardware systems demonstrate features that go beyond the performance of traditional digital electronics. In diffractive free-space optical systems, PNNs at the million-neuron scale have been realized \cite{xu2024large}, with efficient online training for compensating alignment errors and device non-idealities \cite{xue2024fully,zhou2021large}. In coherent PNNs based on MZIs, a single-chip, fully-integrated PNN with six neurons and three cascaded layers achieves ultrafast (410 ps) processing with online training improving the data throughput \cite{bandyopadhyay2024single}. Additionally, a system-on-chip microwave photonic processor using MRR weight banks is developed to solve dynamic radio-frequency (RF) interference \cite{zhang2024system}, showing real-time adaptability enabled by online learning and weight adjustments. We envision that the online training and pruning method presented in this study is also generalizable to other PNNs being actively investigated. For example, the generalized loss function ($\mathscr{L}$) writes as 
\begin{equation}
    \label{Equation: newlossfunction}
    \mathscr{L} = \mathcal{L} + \Sigma_i \gamma_i \Theta_i,
\end{equation}
where $\mathscr{L}$ is given by the sum of the conventional loss function and all the parameter-aware pruning term $\gamma_i \Theta_i$, $\Theta_i$ denotes any hardware parameter related terms to be optimized, such as power consumption, DC electrical control currents/voltages, and modulator biases.

Taking advantage of low-latency photonic processing, online PNN training can benefit applications that necessitate real-time adaptability, including RF interference cancellation \cite{zhang2024system}, fiber nonlinearity compensation \cite{huang2021silicon}, and edge computing \cite{chen2019deep}. For our online training approach implemented on a hybrid PIC-CPU architecture consisting of an $M\times N$ MRR weight bank, the overall latency per training epoch can be expressed as 
\begin{equation}
    \tau_{\mathrm{per\ epoch}} = (MN+1) (\tau_{\mathrm{PIC}} + \tau_{\mathrm{weight \ conf}} + \tau_{\mathrm{ADC}} + \tau_{\mathrm{CPU}} ),
\end{equation}
where $MN+1$ is a factor determined by the perturbation-based algorithm (equals to 7 for our experimental setup), $\tau_{\mathrm{PIC}}$ is the on-chip propagation delay (tens of picoseconds), $\tau_{\mathrm{weight \ conf}}$ is the time needed to thermally configure the weights (on the order of milliseconds), $\tau_{\mathrm{ADC}}$ is the latency induced by ADCs, and $\tau_{\mathrm{CPU}}$ is the time needed by CPU to complete the corresponding digital operations. Potential improvements of the overall latency include the following. First, the factor $(MN+1)$ can be reduced by replacing the perturbation-based optimization method with more efficient training algorithms such as multiplexed gradient descent \cite{mccaughan2023multiplexed}, simultaneous perturbation stochastic approximation \cite{bandyopadhyay2024single}, and fully forward mode training \cite{xue2024fully}. Also, slow thermo-optic effects are unsuitable for online training that requires rapid weight updates \cite{lin2024120}. $\tau_{\mathrm{weight \ conf}}$ can be significantly reduced utilizing faster free-carrier absorption effect \cite{patel2015design} or high-speed electro-optic modulators on thin-film lithium niobate \cite{lin2024120}. Moreover, the latency induced by analog-to-digital conversions and digital electronic processing can be mitigated by exploring co-packaging PIC with FPGAs \cite{zhang2024system,zhong2023lightning}, or even eliminated with all-analog cascaded PNNs \cite{bandyopadhyay2024single}.

\section{Conclusion}
To summarize, we have proposed and demonstrated an online training and pruning method on multi-wavelength PNNs with MRR weight banks that addresses the fundamental issue on scalability and energy efficiency due to MRR resonance variations. We experimentally validate our training framework with an iterative feedback system, and show that superior performances of PNNs can be attained without any software-based pre-training involved. By incorporating the power-aware pruning term into the conventional loss function, our approach significantly optimizes overall power consumption in thermally controlled MRR-based PNNs. This study serves as a fundamental methodology for addressing the chip-to-chip variations in PICs, and represents a significant milestone towards building large-scale, energy-efficient MRR-based integrated analog photonic processors for versatile applications including NNs, LiDAR, RF beamforming, and data interconnects.

\section*{Data Availability}
All data used in this study are available from the corresponding authors upon request.

\section*{Code Availability}
All codes used in this study are available from the corresponding authors upon request.

\backmatter

\bibliography{ref}

\section*{Acknowledgments}

This research is supported by the Office of Naval Research (ONR) (N00014-22-1-2527 P.R.P.), NEC Laboratories America (Princeton E-ffiliates Partnership administered by the Andlinger Center for Energy and the Environment), Princeton's Eric and Wendy Schmidt Transformative Technology Fund, and NJ Health Foundation Award. The devices were fabricated at the Advanced Micro Foundry (AMF) in Singapore through the support of CMC Microsystems. B. J. Shastri acknowledges support from the Natural Sciences and Engineering Research Council of Canada (NSERC). 

\section*{Author contributions}
J.Z., W.Z., and T.X. conceived the ideas. J.Z. performed the simulation, and designed the experiment with support from W.Z. W.Z. developed the experimental photonic setup, including the co-integration of DAC control board and the associated control software. J.Z. conducted the experimental measurements and analyzed the results with support from W.Z. and E.A.D. J.Z. wrote the manuscript with support from L.X., E.A.D., B.J.S., and C.H. P.R.P. supervised the research and contributed to the vision and execution of the experiment. All the authors contributed to the manuscript.

\section*{Competing interests}
The authors declare no competing interests.

\end{document}